# Investigation of charge carrier dynamics in $Ti_3C_2T_x$ MXene for ultrafast photonics applications

Ankita Rawat[1], Nitesh K. Chourasia[1], Saurabh K. Saini[2], Gaurav Rajput[1], Aditya Yadav[2], Ritesh Kumar Chourasia[3], Govind Gupta[2], P. K. Kulriya[1*]

[1]School of Physical Sciences, Jawaharlal Nehru University, New Delhi 110067, India
[2]CSIR-National Physical Laboratory, New Delhi 110012, India
[3] Post-Graduate Department of Physics, Samastipur College (A Constituent College of L.N.M.U.- Darbhanga-846004, Bihar, India), Samastipur-848134, Bihar, India

## Abstract

The rapid advancement of nanomaterials has paved the way for various technological breakthroughs, and MXenes, in particular, have gained substantial attention due to their unique properties such as high conductivity, broad-spectrum absorption strength, and tunable band gap. This article presents the impact of the process parameters on the structural and optical properties of $Ti_3C_2T_x$ MXene for application in ultrafast dynamics. XRD along with Raman spectroscopy studies, confirmed the synthesis of a single phase from their MAX phase $Ti_3AlC_2$. The complete etching of Al and increase in the interplanar distance is also observed on centrifugation at very high speed. The ultrafast transient absorption spectroscopy used to understand the effect of centrifuge speed on the charge carrier dynamics and ultrafast spectrum of MXene displayed that the carrier lifetime is critically influenced by rotation per minute (rpm) *e.g.* faster decay lifetime at 10k rpm than 7k rpm. The electronic relaxation probed using the time-resolved photoluminescence (TRPL) technique exhibits an average decay time ($\tau_{av}$) of 5.13 ns and 5.35 ns at the 7k and 10k rpm, respectively, which confirms that the optical properties of the MXene are strongly affected by the centrifuge speed. The synthesized MXene at 10k rpm typically suggests that radiative processes due to longer decay lifetime and experiences fewer non-radiative losses, resulting in enhanced luminescence properties.

**Keyword**

$Ti_3C_2T_x$ MXene, Process parameters, Ultrafast dynamics, Time-resolved photoluminescence

*Author to whom correspondence should be addressed. Present address: School of Physical Sciences, Jawaharlal Nehru University, New Delhi 110067, India, electronic mail:pkkulriya@mail.jnu.ac.in

**Graphical abstract:**

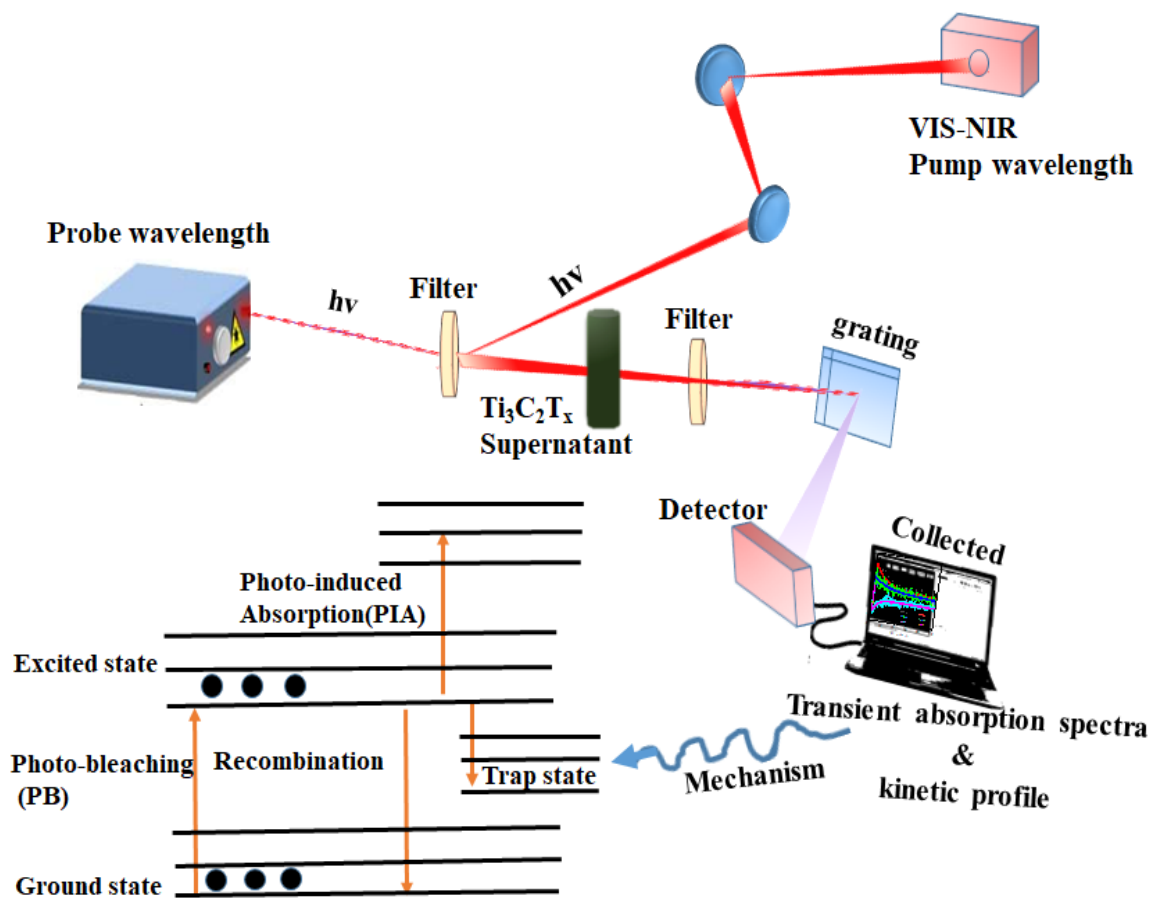

**Introduction**

Recently, two-dimensional materials have received significant attention due to their unique physical and chemical properties compared to their bulk forms.[1] After the isolation of the single-layer graphene in 2004, it has become a reference for all 2D materials and opened the door to the potential of finding even more. Recently, various types of new 2D materials, such as transition metal dichalcogenides (TMDs), hexagonal boron nitride (h-BN), clays, transition metal oxides, etc., have been discovered.[2] The new phase of 2D transition metal carbides, carbonitrides, and nitrides materials known as MXene possesses hydrophilic surfaces with abundant active sites, layered morphology, excellent flexibility, surface terminations controlled tunable metallic or semiconducting behaviour, and good solution processability, making it highly suitable for supercapacitors,[3] sensors,[4] energy storage,[1] optoelectronics,[5] and biomedical applications.[6] MXene has a general formula of $M_{n+1}X_nT_x$, Where n = 1, 2, or 3, and M stands for an early transitional metal, X represents nitrides, carbonitrides, and carbides, while Tx represents various surface termination groups such as -OH, -F, -O.[7] MXene materials can be produced by the removal of A group atoms (especially group 13th and 14th elements like Al, Si, Ga, etc.) from their corresponding MAX phase ($M_{n+1}AX_n$). The "M" atoms of MXene are organized in a packed hexagonal shape with $P6_3/mmc$ symmetry, similar to the MAX phases.[8] Ideally, M atoms engaged with the "X" atoms between the octahedral sites.[9] The M-X bond is a combination of ionic, covalent, and metallic characteristics, while the M-A bond is entirely metallic in nature. In contrast, the bonds that hold the layers together in $M_{n+1}AX_n$ phases are much stronger than the Van der Waals interactions observed in 2D materials like graphene and TMDs. Due to the robust nature of these bonds, it is not possible to mechanically exfoliate $M_{n+1}AX_n$ phases compounds into 2D layers.[10] In the last decade, hundreds of MXenes and their precursor compositions suggested by computational studies and more than 30 MXene phases has been experimentally synthesized. However, there are only few reports on the optimization of the process parameters for the synthesis of impurity free single phase MXene. For example ***Zhang et. al.***[11] synthesized single-phase MXene by exfoliation at different temperatures ranging from 30 °C to 55 °C and delaminated using sonication under Ar flow for one hour, followed by centrifugation at 3500 rpm for one hour. ***Lipatov et al.***[12] optimized the LiF−HCl etching of $Ti_3AlC_2$ for the production of large high-quality MXene flakes having low defect concentration for electronic properties studies. ***Maleski et. al.***[13] research provides insights into how MXene flake size can be controlled by different sonication times and centrifugation and investigates the effect of MXene flake size on the optical and electrochemical properties. The comparative studies on the production of $Ti_3C_2T_x$ MXene from HF etching and LiF-HCl etching showed that the LiF-HCl etched MXene can be delaminated easily by sonication in DI water whereas the HF etched MXene requires additional steps for intercalation with organic molecules to improve its delamination.[14] **Kumar et al.**[15] optimized rotational speed and observed improvement in the energy storage capacity with highest specific capacitance and lowest internal resistance at 10 k rpm due to enhancement of the surface area, capacitance, and interfacial contact of the device. Thus, the properties of the MXene critically depends upon the synthesis parameters such as the centrifugation process, sonication time, molarities of HCl, the chemical composition of the MAX phase, their flake size, and the intercalated species. Therefore, in the present study, the effect of the centrifuge speed on the structural and optical properties of the $Ti_3C_2T_x$ MXene is investigated. Apart from the optimization of the synthesis process of the MXene, the effect of process parameters on the optical properties of $Ti_3C_2T_x$ is also not investigated in detail. ***Jhon et al.***[16] reported MXene-based saturable absorbers for the generation of the ultrafast pulses at the 1.5 µm and 2.0 µm bands. The photoexcited carrier dynamics in $Ti_3C_2T_x$ MXene investigated using ultrafast transient absorption spectroscopy (UTAS) showed a fast relaxation process followed by a slower decay process due to trapping of photoexcited carriers by surface states.[17] Observation of efficient charge separation and long charge carrier lifetime demonstrated the potential of MXene in photovoltaic and photocatalytic applications. ***Colin-Ulloa, et al.***[18] investigated the dynamics of plasmons and free carriers generated by the laser pulse using time-resolved spectroscopy. ***Fu, et al.***[19] probed the dynamics of the light-MXene interaction and flash thermal dissipation using ultrafast transient spectroscopy. To the best of our knowledge, the effect of synthesis parameters on the optical properties of $Ti_3C_2T_x$ has been not investigated using ultrafast transient absorption spectroscopy.

In this article, technologically important impurity free titanium carbide ($Ti_3C_2T_x$) MXene has been synthesized by the LiF/HCl etchants and controlled delamination using varied sonication time and centrifugation processes. The structural, microstructural, compositional, and optical properties of as-prepared MXene has been investigated using XRD, micro-Raman spectroscopy, SEM, EDS, photoluminescence, time-resolved photoluminescence (TRPL) spectroscopy. Furthermore, the excited state dynamics and electronic structure of MXene were also probed using ultrafast transient absorption spectroscopy in femtosecond to nano-second time scales.

## 1. Experimental Details

### 1.1 Materials

Analytical grade reactants such as titanium aluminium carbide $Ti_3AlC_2$ (>90%, <40μm), lithium fluoride powder (300 mesh), hydrochloric acid (HCl, conc. 35%) and deionized water (pH~7) procured from Sigma Aldrich (India), Thermo Fisher Scientific India Pvt. Ltd. and Organo Biotech Laboratories Pvt. Ltd, respectively were used throughout the experiments.

### 1.2 $Ti_3AlC_2$ etched through LiF/HCl

Single-phase impurity free MXene has been synthesized using wet chemical method. Figure 1 provides an overview of the steps involved in the synthesis of MXene. Firstly, Al atoms were removed from their precursor MAX phase $Ti_3AlC_2$ through selective etching with LiF/HCl etchant. The etchant contained 0.8g of LiF dissolved in 10mL of aqueous hydrochloric acid (9 mol L$^{-1}$) solution, and after 30 minutes of stirring, 0.5g of $Ti_3AlC_2$ MAX phase powder was gradually added to the reaction mixture within the course of 15 minutes. After that, a black dispersion was formed after being stirred at room temperature for 24 hrs. The raw products were cleaned many times with deionized water and centrifugation at 4,500 rpm until the pH reached nearly to six. The high-yield precipitated MXene, also known as the sediment, is then obtained after multiple washes at 4,500 rpm centrifugation, as illustrated in Figure 1. The recovered sediment is then mixed with the DI water, sonicated for 2 hrs, then centrifuged at 7000 (7k) rpm for 30 minutes. The obtained dark greenish supernatant is named as S1. To further improve the quality of synthesised MXene, the sonication time is increased to 2 hrs followed by centrifugation higher speed of at 10,000 (10k) rpm for 30 minute. The obtained light greenish supernatant is named as S2. This helps in the dispersion of the sheets into individual layers and the formation of the impurity free MXene. In order to obtain completely exfoliated MXene, a series of repeated centrifugations at progressively higher speeds with higher sonication time were also performed. Finally, the $Ti_3C_2T_x$ MXene sediment was dried in a hot vacuum oven and kept under the $N_2$ environment. The S1 and S2 supernatant were used for further characterization.

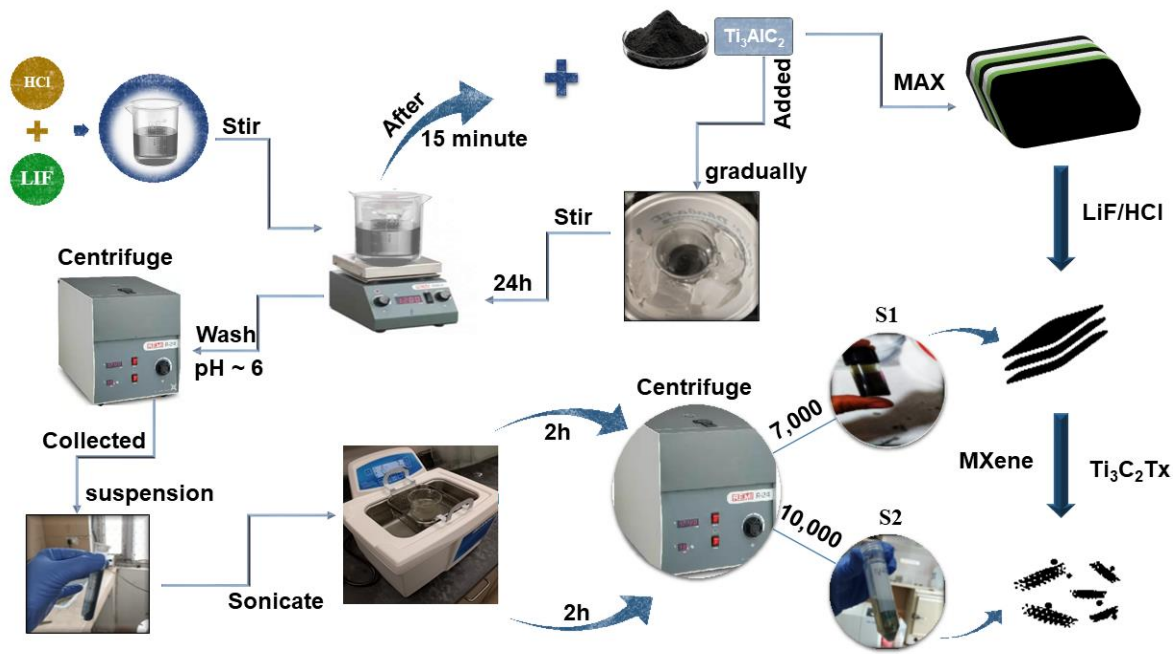

**Figure: 1** The process for synthesis of impurity free single phase $Ti_3C_2T_x$ MXene.

## 2. Materials Characterization

The diffraction pattern of precursor $Ti_3AlC_2$ was recorded through the X-ray diffractometer whereas the grazing incidence X-ray diffraction (GIXRD) pattern of $Ti_3C_2T_x$ MXene supernatant S1 and S2 were recorded by using a PANalyticalX'Pert PRO diffractometer equipped with a Cu-$K_α$ ($λ=$ 0.15406 nm) X-ray source in the range of 5° to 60° with a step size of 0.02° and scan time of 5s at a grazing angle of 1°. Raman spectra were recorded on a Raman spectrometer (WITec alpha 300 RA), using 532.5 nm laser excitation. Surface morphology and elemental analysis of $Ti_3C_2T_x$ MXene were characterized by using scanning electron microscopy (SEM) and energy dispersive spectroscopy (EDS) suppled by ZEISS EVO 40. The optical features were studied using a UV-Vis spectrophotometer (HITACHI U-3900) and photoluminescence (FLS980 D2D2), Edinburg, double monochromatic system Xenon lamp & TRPL 266 laser spectroscopy. Ultrafast transient absorption spectroscopy (UFTAS) technique utilizes a Ti: Sapphire laser, such as the Coherent Micra (oscillator) and Legend USP (amplifier). The oscillator produces mode-locked Gaussian-shaped pulses with a pulse width of 35-45 fs and a repetition rate of 80 MHz at 800 nm with a tuning cpability form 780-810nm. The amplifier provides a pulse width of 35-45 fs with a repetion rate of 1 kHz and centerd at 800 nm with an outuput average power of 3-4 mJ. The output of the amplifier beam was then split into two parts using beam splitters. One portion was directed to the operational parametric amplifier (OPA) to generate a pump, while the other portion was delayed and enters the spectrometer (Helios from Ultrafast systems Inc.) after passing through a sapphire disk to produce a white light continuum (WLC) that serves as the probe beam[20], which is used to determine the kinetics and carrier dynamics studies. Fitting is accomplished using the surface Xplorer analysis and Origin 9.1 softwares .

## 3. Results and Discussions

### 3.1. Structural and microstructural properties:

The X-ray diffraction (XRD) along with the Raman spectroscopy technique, is used to probe the structural properties $Ti_3C_2T_x$ MXene and its supernatants. Figure 1 shows the XRD pattern of (a) $Ti_3AlC_2$ powder, $Ti_3C_2T_x$ MXene (before sonication), $Ti_3C_2T_x$ MXene supernatant S1 and S2 (after sonication) and (b) Raman spectra of $Ti_3C_2T_x$ MXene supernatant S1 and S2. It can be seen that the diffraction peak corresponding to MAX phase of $Ti_3AlC_2$ are appeared at $2\theta \approx 9.8°$ and assigned as (002) plane and new broadened peak are also appeared at lower angle $2\theta \approx 6.5°$ after etching in HCl + LiF solution for 24 hrs, followed by sonication and centrifugation at 7k rpm. In addition, small sharp peak of Al (104) at $2\theta \approx 38.6°$ is also observed. The presence of these peaks are clear evidence that MAX phase is not completely etched at 7k rpm. When centrifugation rpm is increased from 7k to 10k then the peaks corresponding to MAX phase $Ti_3AlC_2$ (002) peak and Al (104) peak get disappeared and two distinct peaks at $2\theta \approx 6.1°$, and an insignificant amount of Al at 38.6°, assigned to the (002) and (104) planes of $Ti_3C_2T_x$ MXene, respectively; are observed. The lower angle shifting of MXene is a direct evidence for synthesis of the single phase $Ti_3C_2T_x$ MXene which is also consistent with previous studies on MXene.[15,21,22] It may be noted that the position of the (002) diffraction peak is shifted from 6.5° (for S1) to 6.1° (for S2) on the sonication and centrifugation of the $Ti_3C_2T_x$ sample indicating an increase in the interplanar distance from 9.12 Å (for $Ti_3AlC_2$) to 13.58 Å (for S1) and 14.47 Å (for S2), respectively. The calculated values of the lattice parameters are 18.03 Å, 27.17 Å and 28.95 Å for $Ti_3AlC_2$, S1 and S2 samples, respectively which is consistent with earlier reported data on $Ti_3C_2T_x$ crystal structures.[23,24] The structural parameters of as-synthesized MXene are shown in Table 1.

| Sample | Plane | 2θ | d-spacing | c |
|---|---|---|---|---|
| MAX ($Ti_3AlC_2$) | (002) | 9.8 ° | 9.12 Å | 18.03 Å |
| MXene (Before Sonication) | (002) | 7.4 ° | 11.93 Å | 23.87 Å |
| S1 | (002) | 6.5° | 13.58 Å | 27.17 Å |
| S2 | (002) | 6.1° | 14.47 Å | 28.95 Å |

**Table-1** Different structural parameters of as-synthesized MXene.

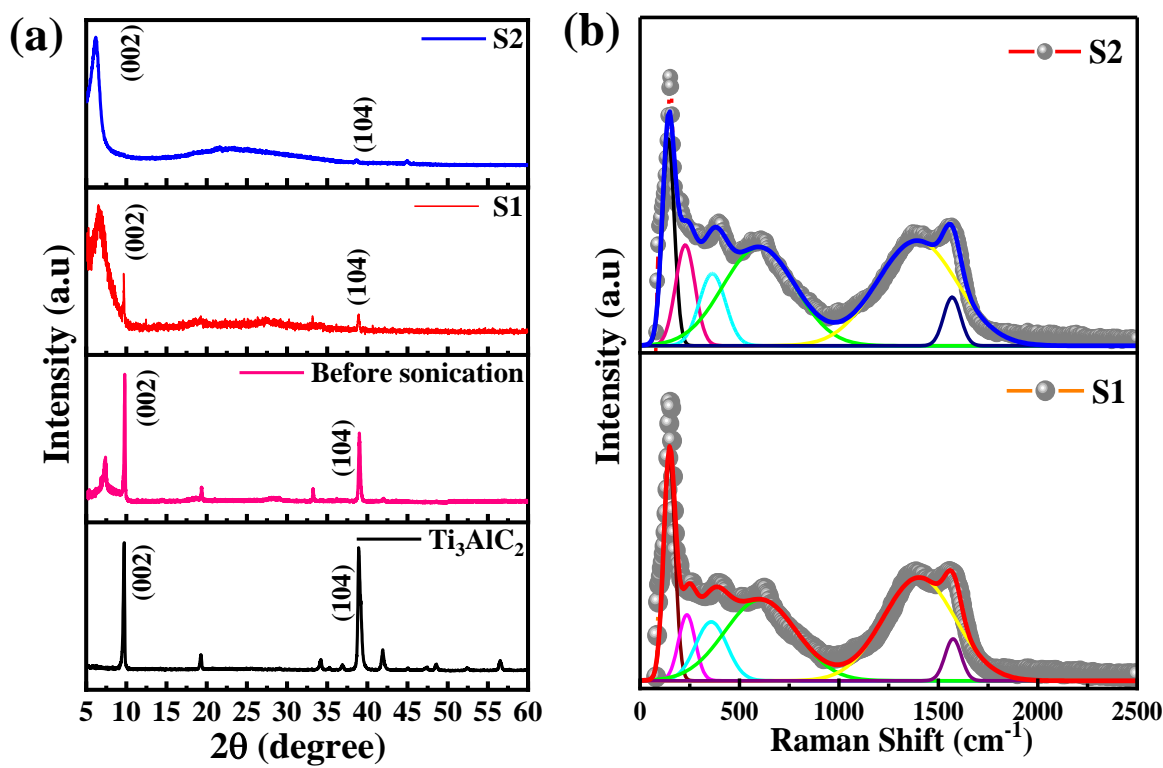

**Figure 2:** XRD pattern of **(a)** Ti$_3$AlC$_2$ powder, Ti$_3$C$_2$T$_x$ MXene (before sonication), Ti$_3$C$_2$T$_x$ MXene supernatant S1 and S2 (after sonication) and **(b)** Raman spectra of Ti$_3$C$_2$T$_x$ MXene supernatant S1 and S2.

The Raman spectroscopy that provides information about the short-range structural ordering is also used for characterizing the structural and vibrational properties of Ti$_3$C$_2$T$_x$ MXene. The Raman spectra of the initial MAX phase Ti$_3$AlC$_2$ has been shown in supporting information [Figure 4]. The Raman peaks of Ti$_3$AlC$_2$ located around the wavenumber of 198 cm$^{-1}$, 268 cm$^{-1}$ and 621 cm$^{-1}$, respectively are observed due to the shear and longitudinal oscillations of the Ti and the Al atoms.[25] In particular, the peak located at 268 cm$^{-1}$ is associated with vibrations of the Al as reported earlier[26]. The absence of Al peak in the Raman spectrum [Figure 2b] is a direct outcome of the extensive etching of Al atoms that leads to the formation of MXene structure.[27] The Raman spectrum of Ti$_3$C$_2$T$_x$ MXene shows several prominent peaks due to its different vibrational modes. The most intense peak appeared at 145 cm$^{-1}$ is attributed to the doubly degenerated E$_g$ modes, which corresponds to the vibration of the Ti atoms in the Ti$_3$C$_2$T$_x$ layers.[28] The other peaks are observed at 235 cm$^{-1}$, 357 cm$^{-1}$ and 606 cm$^{-1}$ for S1 and 227 cm$^{-1}$, 363 cm$^{-1}$ and 597 cm$^{-1}$ for S2, respectively. The peaks that appeared at ~ 230 cm$^{-1}$ and ~ 360 cm$^{-1}$ are associated with the Ti-C vibrations and O atoms vibrations, whereas the peak observed at ~ 600 cm$^{-1}$ is related to the E$_g$ vibrations of the carbon in the Ti$_3$C$_2$T$_x$ MXene that have terminal groups.[15,29] Two additional bands appeared at 1392 cm$^{-1}$ and 1565 cm$^{-1}$ for S1, 1371 cm$^{-1}$ and

1548 cm$^{-1}$ for S2 are assigned as D and G bands, respectively.[28] Here, D band represents the vacancies and other structural defects in the Ti$_3$C$_2$T$_x$ MXene where as G band exhibited in-plane stretching vibration of carbon in MXene layers.[30] The observation of the D and G bands between 1000-1800 cm$^{-1}$ indicates that S1 and S2 have almost similar structure.[15,28] Furthermore, observation of the lower intensity D band as compared to G band is also in good agreement with the previous reported literature.[31]

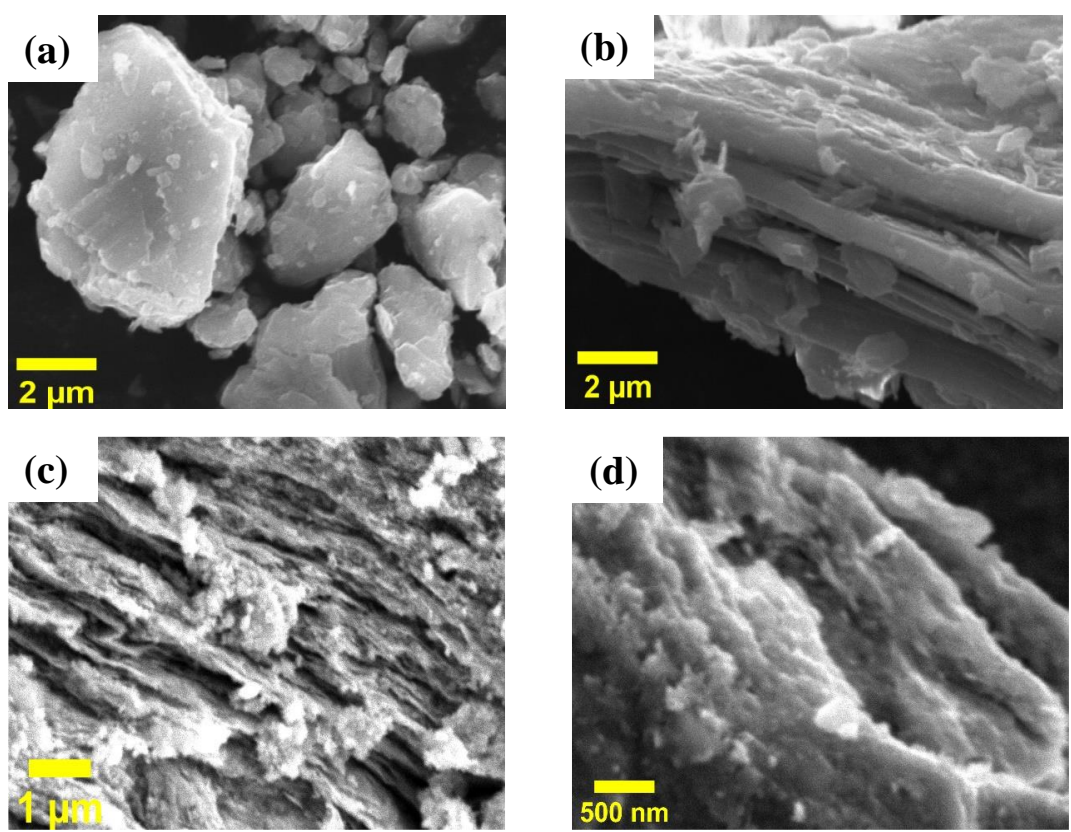

**Figure 3:** Surface morphologies of **(a)** Ti$_3$AlC$_2$ MAX phase, (b) Ti$_3$C$_2$T$_x$ MXene sediment, (c) Ti$_3$C$_2$T$_x$ MXene supernatant S1 and (d) S2 after sonication.

The composition and surface morphologies of MAX phase as well as as-synthesized MXene were investigated using energy dispersive spectroscopy (EDS) and scanning electron microscopy (SEM), respectively. The micro sheets of the Ti$_3$AlC$_2$ MAX phase exhibit a modest laminated pattern on their side surfaces [Figure 3a].[32] After LiF/HCl etching, the morphologies of the MXene powder, S1, and S2 MXene supernatants as have completely developed into an accordion-like laminated structure with fabric-like layered features [Figure 3(b-d)].[33] This demonstrates the successful exfoliation of the MAX phase into multilayer MXene.

The elemental distributions for Ti$_3$AlC$_2$ MAX phase and Ti$_3$C$_2$Tx MXenes (S1 and S2) determined by EDX detector (see supplementary information, Figure 1, 2 and 3) shows that the laminated $Ti_3C_2$ primarily consists of Ti and C, along with a small amount of O, F and Cl elements.[22] The existence of F and O may be related to the terminal groups, while the presence of Ti and C is predictable. However, the identification of low quantities of Al and Cl are most likely attributable to an incomplete etching process or residue that was not washed away, which leads to impurities or the formation of a terminal group.[31] The concentration of different elements are shown in the supplementary information (see figure 1, 2 and 3) for $Ti_3AlC_2$ MAX phase, S1, and S2, respectively. It was found that the majority of the Al is removed during the etching process, as S1 has 1.29 atomic percentage of Al and S2 has 0.20 atomic percentage, in comparison to the $Ti_3AlC_2$ MAX phase, which has 20.85 atomic percentage of Al. It is important to note that S2 contains very small concentration of Al element and almost a single phase of $Ti_3C_2T_x$ MXene. Through EDX, the presence of Ti, C, and O was confirmed as major elements of $Ti_3C_2T_x$ composition in both the supernatant S1 and S2. In addition, the traces of F, and Cl were also found as elements of the terminal groups in both MXene supernatant S1 and S2, respectively. The surface area of MXene mostly adsorbs hydroxyl-OH, fluorine-F, and oxygen-O during wet-etching processing[34].

## 3.2. Optical Properties

The optical properties and electronic structure of the $Ti_3C_2T_x$ MXene has been probed using photoluminescence spectroscopy, time-resolved photoluminescence (TRPL) spectroscopy and ultrafast transient absorption spectroscopy technique. According to *Zhang et al.*[35], band gap of $Ti_3C_2T_x$ MXene powder is just about 0.1 eV, which is not large enough to emit light in the visible or near-infrared region. *Xue et al.*[36] suggested that the luminescence properties of $Ti_3C_2T_x$ MXene depend on size and surface defects which leads to a strong quantum confinement effect due to its small lateral size. Thus, a reduction in the size of the MXene leads to increase in its bandgap as well as impels the efficiency for emitting light. Apart from that, the luminescence in $Ti_3C_2T_x$ MXene also arises from surface defects, which create energy states within the bandgap of MXene that assist in light emission when it is excited. Surface defects can be caused by various factors, such as oxidation, surface functionalization, or structural defects.[37] When defects of the supernatant of $Ti_3C_2T_x$ MXene create sites where oxygen can be adsorbed onto its surface/solution and react with Ti atoms, resulting in the formation of $TiO_x$.[35] Thus, surface defects plays important role in the determination of the optical properties of the $Ti_3C_2T_x$ MXene. The photoluminescence spectroscopy measurement was performed in the wavelength range of 390 nm to 800 nm with an excitation wavelength of 374 nm at room temperature.

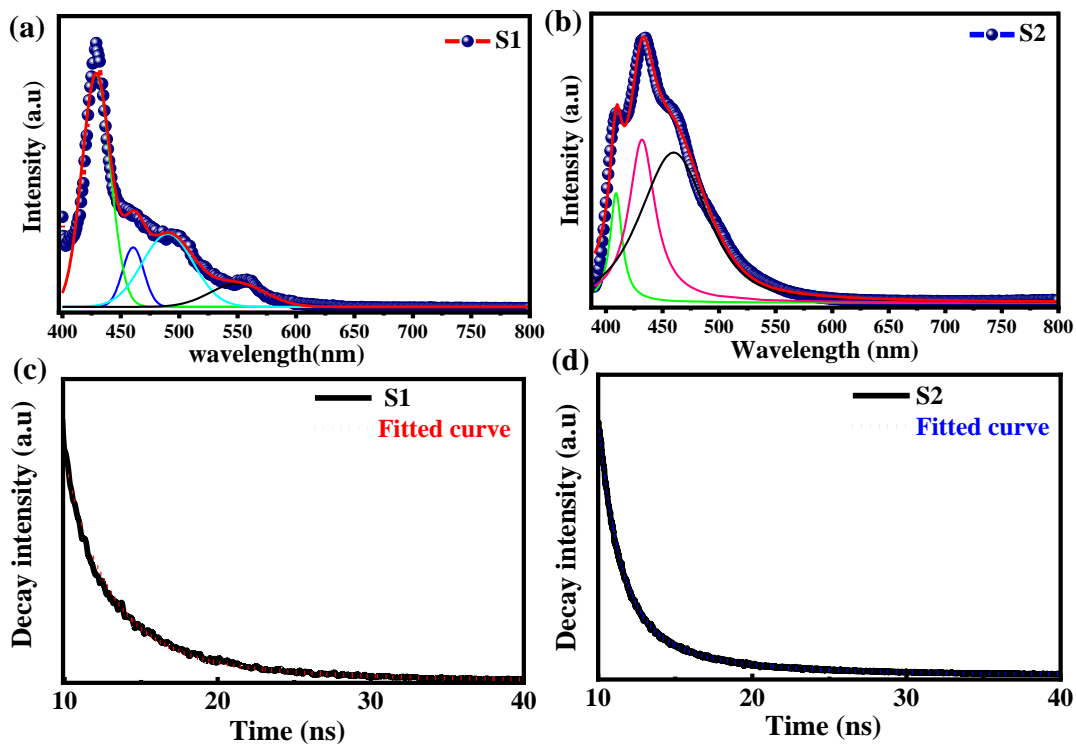

**Figure 4:** Emission spectra of $Ti_3C_2T_x$ MXene at **(a)** S1 and, **(b)** S2. Time-resolved photoluminescence (TRPL) spectra of $Ti_3C_2T_x$ MXene supernatant, **(c)** S2 and **(d)** S1.

The photoluminescence spectra of MXene supernatant (S1 & S2) is shown in Figure 4 (a-b). The two prominent emission peaks of MXene supernatant (S1 & S2) were positioned at 428 nm and 434 nm. On the deconvolution fitting of broad PL spectra, it revealed the presence of three well resolved shallow trap levels at a wavelength of 460 nm, 490 nm and 550 nm for supernatant S1, and two peaks at the 408 nm, and 460 nm for supernatant S2.

The peaks appeared at the high-energy peaks (428 nm for S1 and 434 nm for S2) to band edge luminescence are assigned to the $Ti_3C_2T_x$ MXene supernatant, while the lower-energy peaks are induced by the presence of the oxygen vacancies and surface functional group, which arises due to different experimental conditions such as temperature, sonication, solvent and aging time during synthesis.[38] Also, the presence of oxygen vacancy states in the band gap of the MXene supernatant band structure can contribute to the appearance of the visible light absorption.[36] The absorption spectra of MXene supernatant S1 and S2 has been shown in the supplementary information [Figure 5].

To analyse the dynamical process such as excitation, relaxation, and recombination, time-resolved photoluminescence (TRPL) spectroscopy of the $Ti_3C_2T_x$ MXene has been performed. The TRPL spectra for $Ti_3C_2T_x$ MXene synthesized at 7k rpm and 10k rpm is shown in Figure 4

(c, d). The decay time ($\tau$) of the photoexcited charge carrier was estimated through bi-exponential TRPL decay fitting by using equation (1),

$$I(t)=A+B_1\exp(-t/\tau_1)+B_2\exp(-t/\tau_2)+B_3\exp(-t/\tau_3)+\ldots\ldots..B_i\exp(-t/\tau_i) \qquad (1)$$

Where I (t), A, and $B_i$ are luminescence intensity, constant, and PL intensity of ith components (t = 0), respectively, $\tau_1$ and $\tau_2$ are short and fast decay time constants, respectively. The calculated values of decay parameters of S1 were $\tau_1$ (ns) = 2.71 ± 0.03, $\tau_2$ (ns) = 9.50 ± 0.00, $B_1$ = 1210.25 ± 11.84, $B_2$ = 220.87 ± 3.38, and $\chi^2$ = 0.990. These parameters were also calculated for MXene supernatant S2, which were $\tau_1$ (ns) = 1.34 ± 0.01, $\tau_2$ (ns) = 4.10 ± 0.09, $\tau_3$ (ns) = 13.1 ± 0.1, $B_1$ = 3151.92 ± 30.42, $B_2$ = 892.05 ± 28.98, $B_3$ = 187.52 ± 6.70, and $\chi^2$ = 1.044. It may be noted that decay time constants $\tau_1$, $\tau_2$ and $\tau_3$ are related to the slow and fast decay process, which is also reflected in the calculated parameters where the value of $\tau_1 < \tau_2$ for S1, and $\tau_1 < \tau_2 < \tau_3$ for S2. Fast and slow decay processes are related to defect-assisted radiative/non-radiative recombination. Luminescence Lifetimes are measures of the amount of time, in terms of its average decay time ($\tau_{av}$), that a luminous material remains in an excited state before falling back to its ground state. By analyzing the decay curves in the TRPL spectra, the average decay time ($\tau_{av}$) was estimated to be 5.13 ns for S1 and 5.35 ns for S2 using equation (2).[38]

$$\tau_{av} = B_1\tau_1^2 + B_2\tau_2^2 + \ldots\ldots\ldots.B_i\tau_n^2 \Big/ B_1\tau_1 + B_2\tau_2 + \ldots\ldots\ldots+B_i\tau_n \qquad (2)$$

A slower rate of luminous signal decay can be inferred from a TRPL spectrum of supernatant S2 that exhibits a greater lifespan decay. It gives the impression that the luminous material stays in its excited state for a longer period of time before falling back to its ground state. This information has the potential to shed significant light on the processes, both radiative and non-radiative, that are taking place in the material. Non-radiative processes include the dissipation of energy through mechanisms such as vibrational relaxation or quenching, whereas radiative processes result in the emission of light from the system. Supernatant S2 often indicates that radiative processes are taking place due to a longer lifetime, and as a result, it experiences fewer non-radiative losses, which results in increased luminescence properties.

Further, the transient absorption study of MXene supernatant material is carried out through ultrafast transient absorption spectroscopy. The ultrafast spectrum is obtained from MXene supernatant S1, and S2 synthesized at two different rpm (7k and 10k). Moreover, MXene materials were excited through a 350 nm pump wavelength at the average pump power is 1mW, and the supernatant S1 spectrum is recorded in a broad visible to near-infrared probe range (450 nm to 1450), as shown in Figure 5 (a, b). The transient absorption (TA) spectra of S1 MXene are quite different from S2 MXene spectra. The photo-induced absorption (PIA) and ground state bleaching (GSB) are clearly observed in the visible probe range [Figure 5a]. This could lead to the formation of trap-states due to bigger particle size, agglomeration, and not completely etching the MAX phase of the MXene at 7k rpm (S1). In the case of the NIR probing, a broad PIA in the entire range 800 nm to 1450 nm is also detected due to electron-phonon scattering [Figure 5b]. The increasing PIA signal with the increase in probe wavelength indicates the enhancement of the charge carrier density in the excited state.

Furthermore, the MXene supernatant S2 is excited at the same pump wavelength, and spectra are also recorded in both visible and near-infrared probe ranges between 450 nm and 1450 nm, as shown in Figure 5 (c, d).

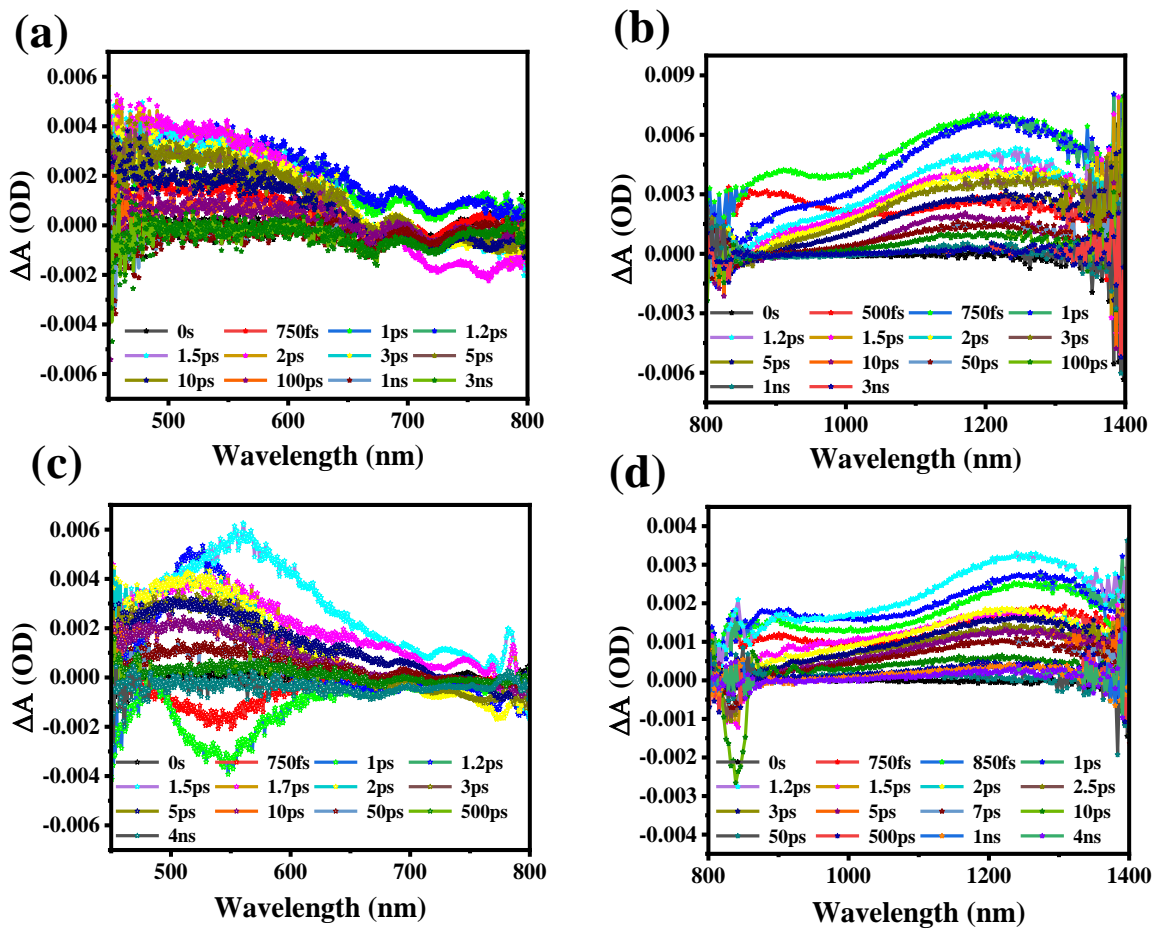

**Figure 5:** Ultrafast spectrum of $Ti_3C_2T_x$ MXene supernatant S1 & S2 (a, b), transient absorption spectra at 7k rpm (S1) and (c, d) transient absorption spectra of MXene supernatant at 10k rpm (S2) excited at 350 nm pump wavelength and data will record in both visible and NIR range respectively.

Observation of a widespread photo-induced absorption (PIA) in the visible probe spectrum (450 nm to 800 nm) clearly exhibits an electron-phonon interaction in excited states. Due to excited charge carriers shifting from the valence to conduction bands, ground state bleaching (GSB) between 750 fs and 1ps probe delay periodic also seen in this spectrum. This displays highly efficient charge carrier dynamics due to a reduction in micro-strain and lattice relaxation [Figure 5c]. The spectrum is also recorded in the broad NIR probe regime between 800 nm and 1450 nm, where PIA was observed across the probe range. At higher probe wavelengths, the transient absorption increases with the maximum absorption observed at 1230 nm [Figure 5d]. The broad PIA is observed due to the electron-phonon scattering in the excited states of the conduction band.

Further, the ultrafast carrier kinetic profile of MXene supernatant at 7k rpm (S1) in both visible and near-infrared probe regimes, as shown in Figure 6 (a, b).

| Table 2 : | | | | | |
|---|---|---|---|---|---|
| Sample @Pump Wavelength (nm) | Wavelength (nm) | $t_1$ (ps) | $a_1$ (%) | $t_2$ (ps) | $a_2$ (%) |
| S1 (7k rpm) 350 nm, Visible regime | 500 | 7.56 | 64.4 | 135 | 35.6 |
| | 600 | 10.3 | 59.6 | 170 | 40.4 |
| | 755 | 1.78 | 93.2 | 3.15 | 6.8 |
| S1 (7k rpm) 350 nm, NIR regime | 900 | 1.17 | 95.9 | 1.67 | 4.09 |
| | 986 | 1.2 | 87.5 | 136 | 12.5 |
| | 1200 | 1.39 | 78.4 | 46.8 | 21.6 |
| S2 (10k rpm) 350 nm, Visible regime | 525 | 0.23 | 65.8 | 12 | 34.2 |
| | 600 | 0.186 | 83.3 | 15.1 | 15.7 |
| | 728 | 0.187 | 85.1 | 1.93 | 14.9 |
| S2 (10k rpm) 350 nm, NIR regime | 900 | 0.62 | 93.8 | 118 | 6.2 |
| | 1154 | 1.18 | 63.2 | 36 | 36.7 |
| | 1253 | 1.12 | 74.2 | 33.3 | 25.8 |

**Table 2:** The fitting parameters for the kinetics profile of MXene (S1 & S2) for both visible and near-infrared probe regimes.

The kinetic profiles of both probes are fitted through the model fitted by equation 3 in the "Surface Xplorer" software and calculated the fitting parameters such as $t_1$, $t_2$, and $a_1$, $a_2$ which are listed in **Table 2.** The times $t_1$, and $t_2$, show the initial rise time and charge recombination time, respectively.

The following function enables fitting a kinetic trace with a sum of convoluted exponentials at the selected wavelength:

$$S(t) = e^{-(\frac{t-t_0}{t_p})^2} * \sum i A_i e^{-((t-t_0)/t_i)} \tag{3}$$

$A_i$ and $t_i$ are the amplitudes and decay times, respectively, where $t_p$ is the instrument reaction time, and $t_0$ is time zero. The rise and recombination times are enhanced due to agglomeration and the fact that the MAX phase in the MXene was not etched completely at 7k rpm (S1). This incomplete etching leads to electron-photon coupling, which affects the behavior of the charge carriers [Table 2]. In addition, PIA and GSB both are also observed in the visible probe (450 nm to 800 nm).

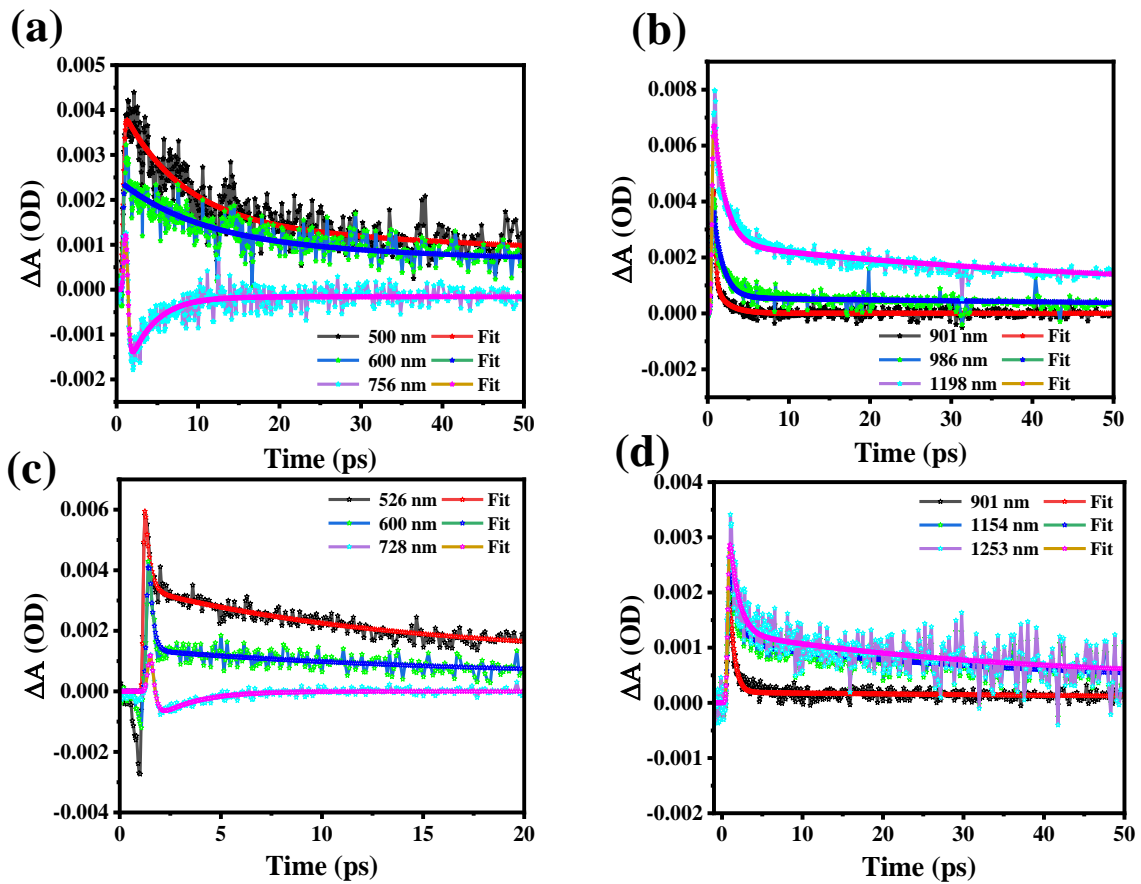

**Figure 6.** Transient absorption spectrum of MXene supernatant S1 & S2 **(a, b)** kinetic profiles of MXene supernatant at 7k rpm (S1) and **(c, d)** kinetic profiles of MXene supernatant at 10k rpm (S2) excited at 350 nm pump wavelength and data recorded in both the visible and NIR range.

The corresponding kinetic profile of MXene supernatant at 10k rpm (S2) in both visible and near-infrared probe regimes, as shown in Figure 6 (c, d). The kinetic profile in the visible range of MXene supernatant S2 shows the hot charge carrier interaction with phonon in the excited states as well as the relaxation of hot charge carriers from the highly excited states to quasi-thermal equilibrium states, which gives rise to state filling of MXene, as shown in [Figure 6c]. In addition, the rising time of carriers is depicted in the sub-femtoseconds due to intraband relaxation and electron-electron scattering in the excited states. Moreover, the kinetic profiles of the NIR probe show photo-induced absorption with a very short carrier rise time, as shown in [Figure 6d]. The charge recombination time is increased due to electron-phonon coupling. In comparison, Kinetic profiles between VIS and NIR probe regime of supernatants S1 and S2 suggests that supernatant S1 exhibits a more compact behaviour in terms of rise and recombination times of charge carriers due to agglomeration and incomplete etching of the MAX phase in the MXene.

**Conclusion**

$Ti_3C_2T_x$ MXene has been successfully synthesised by etching Al from MAX phase by LiF/HCl route. The effect of centrifugation speed on the quality and the property of MXene has shown that MXene synthesised at 10k rpm is better than synthesised at 7k rpm in terms of material quality and device perspectives. The MXene synthesized at 10k rpm also exhibits faster and efficient charge carrier capabilities, which is required for high-speed photonic devices. Using the time-resolved photoluminescence techniques, the electronic relaxation (average decay time $\tau_{av}$) was estimated at 5.13 ns and 5.35 ns for 7k and 10k rpm, respectively. Supernatant S2 (10k rpm) typically suggests that radiative processes due to longer decay lifetime and experiences fewer non-radiative losses, resulting in enhanced luminescence properties. This finding indicates that the optical properties of the MXene are strongly influenced by the process parameters used in the synthesis of the material.

**Acknowledgment**

This work was supported by "Science and Engineering Research Board," India (PDF/2021/001490). Nitesh K. Chourasia acknowledges "Science and Engineering Research

Board'' for National Post-Doctoral Fellowship (PDF/2021/001490). Ankita Rawat thanks CSIR-UGC NET with File No. 09/263(1233)/2020-EMR-I for providing Junior Research Fellowship (JRF). The authors are very thankful to Dr. Mahesh Kumar for providing ultrafast facilities in the CSIR-NPL, New Delhi, India.